\documentclass[12pt]{article}
\usepackage[utf8]{inputenc}
\usepackage{amssymb}
\usepackage{graphicx}
\usepackage{xcolor}
\usepackage{lineno}
\usepackage{openwork}
\usepackage{setspace}
\fancypagestyle{plain}{%
  \fancyhf{}
  \fancyfoot[R]{\textcolor{black}{\thepage}}
  
}

\title{Evaluating PhaseNet on Teleseismic Data with MsPASS}
\author[1]{Jinxin Ma}
\author[1]{Yinzhi Wang}
\author[2]{Gary L. Pavlis}
\author[1]{Chenbo Yin}
\affil[1]{Texas Advanced Computing Center, The University of Texas at Austin}
\affil[2]{Department of Earth and Atmospheric Sciences, Indiana University, Bloomington, IN 47405}
\date{}
\newcommand{\alttext}[1]{\par\vspace{0.35\baselineskip}{\small\noindent\raggedright\textit{Alt text:} #1\par}}

\begin{document}


\maketitle






\begin{abstract}
Numerous studies have shown that the machine-learning picker PhaseNet produces accurate P and S picks on local earthquake signals, but its performance can degrade sharply on teleseismic signals. To address this limitation, we present a reproducible MsPASS workflow that (i) enables scalable data preparation and management for large seismic archives and (ii) supports standardized PhaseNet training and inference. We assembled a control dataset of 1.6 million waveforms linked to teleseismic P-wave picks made by analysts at the USArray Array Network Facility (ANF). The control dataset confirms that the PhaseNet model trained on regional signals performs poorly on these data. We then trained PhaseNet from scratch on the training split of the ANF control dataset and evaluated it on a non-overlapping held-out test split, increasing P-pick recall by 741.5\% and yielding 683.9\% more picks within a 0.1~s residual window.
We also evaluated PhaseNet across different model sizes on both CPUs and GPUs. Increasing the model size by about 120 times improved precision and recall by 15.6\% and 23.2\%, respectively. However, the scaled model reduced inference throughput by 87.2\% on an NVIDIA A100 GPU and by 97.3\% on a 128-core high-performance CPU node. These results indicate that scaling PhaseNet is more practical on GPUs than on CPUs, and that simply enlarging the model is not an efficient way to achieve large accuracy gains. \\
\end{abstract}

\section{Introduction}
Accurate seismic phase picks are essential for earthquake monitoring because they underpin network catalogs. As the number of recording channels has grown, manual picking has become increasingly difficult and expensive for network operations. Human analysts are highly skilled, but they are limited in number and can process only a finite amount of data each day. \textcite{johnson1979cedar} showed that detectors based on the standard STA/LTA (short-term average divided by long-term average) approach formed the backbone of real-time seismic processing from the earliest days of digital data acquisition in seismology. A number of variants based on more sophisticated signal-processing methods have also been proposed, with mixed success, including those of \textcite{allen1978automatic} and \textcite{saragiotis2002pai}.

The fundamental problem faced by all seismic detectors is that both real seismic signals and real background noise are highly complex. As a result, every detector makes mistakes. In practice, these errors are reduced through parameter tuning and through decision rules that automatically discard inconsistent data. The most common example is event association: a detection is considered valid only when a set of potential detections is consistent with seismic wave propagation for one or more phases.

In the past decade, deep-learning models have emerged as an alternative to conventional pickers. \textcite{zhu2019phasenet} introduced PhaseNet, which is now widely used in both method-development and comparative benchmark studies such as \textcite{munchmeyer2022picker}. It takes three-component waveform windows as input and outputs sample-by-sample probabilities for P, S, and noise, from which arrival-time picks are derived. Other influential approaches include generalized phase detection (GPD) by \textcite{ross2018generalized},
EQTransformer (multi-task detection and picking with attention) by \textcite{mousavi2020earthquake}, and two-stage pickers such as DeepPhasePick by \textcite{soto2021deepphasepick}.

A key challenge for deep-learning pickers is generalization to waveforms that differ from those seen during training. Waveform characteristics change with source-receiver distance, frequency content, and noise conditions. In particular, teleseismic arrivals differ systematically from local and regional arrivals: they typically have lower dominant frequencies and envelope shapes that depend on source depth and source complexity. It is therefore not surprising that models trained on local and regional signals often perform poorly on teleseismic data.

While \textcite{munchmeyer2022picker} showed that models trained on regional data often transfer poorly to teleseismic recordings, the USArray waveform archive and the associated ANF bulletin still provide a valuable setting for studying this problem. USArray was a continental-scale seismic experiment centered on the Transportable Array, a rolling network of more than 400 broadband stations deployed in a quasi-uniform grid with roughly 70~km spacing across much of the United States. \textcite{doi.org/10.7914/sn/ta} documented that its waveform data are openly archived through EarthScope/IRIS, making the dataset widely accessible for subsequent methodological studies. \textcite{trabant2012data} described the associated Array Network Facility (ANF) bulletin, which provides phase-arrival and event information derived from these recordings, with analyst review for a substantial portion of the earlier archive. \textcite{pavlis2010array} further showed that USArray recordings support efficient large-scale array processing of teleseismic body waves, reinforcing its value as a practical testbed for teleseismic evaluation.


It is unclear whether scaling up PhaseNet is worthwhile, or what hardware is needed to do so efficiently. \textcite{kaplan2020scaling} showed in the context of large language models that increasing model size can yield dramatic performance gains. However, scaling laws depend strongly on model architecture, data volume, and task complexity. For a relatively compact model like PhaseNet, it is not obvious that simply increasing model size will yield proportional accuracy gains.

To the best of our knowledge, no prior study has evaluated the accuracy and efficiency of PhaseNet across different model sizes and hardware configurations on teleseismic signals. In the original PhaseNet study, \textcite{zhu2019phasenet} focused on accuracy metrics including residual histograms and standard classification metrics for P/S picks. \textcite{naoi2024neural} introduced a PhaseNet-based model (PhaseNetWC) with increased convolutional channel capacity and demonstrated improved performance on Japanese datasets. However, these accuracy-focused comparisons still rarely characterize the computational behavior of these models. \textcite{yu2023benchmark} evaluated multiple deep-learning phase pickers on the regional DiTing dataset and included CPU/GPU performance measurements, but their analysis did not address teleseismic deployment or consider different PhaseNet model sizes.

In this paper, we evaluate the phase-picking accuracy and efficiency of PhaseNet on teleseismic waveforms from USArray. This study makes three practical contributions:
\begin{enumerate}
    \item A reproducible MsPASS workflow for large-scale seismic dataset building, training, and inference. Using this workflow, we process 125.5 million USArray waveforms and compile a machine-learning dataset of 1.6 million waveforms linked to analyst picks and stored in ObsPy trace format.
    \item An evaluation on USArray teleseismic waveforms showing that the standard PhaseNet model trained on regional signals generalizes poorly, and that training on teleseismic data recovers performance, increasing recall by 741.5\%, F1 by 383.1\%, and the number of picks within a 0.1~s residual window by 683.9\%, while precision decreases by 33.2\%.
    \item A benchmark showing that increasing PhaseNet size does not guarantee proportional accuracy gains, and that scaling PhaseNet is more practical on GPUs than on CPUs. Specifically, a roughly 120-fold increase in model size improves precision and recall by 15.6\% and 23.2\%, while reducing NVIDIA A100 GPU inference throughput by 87.2\% and CPU inference throughput by 97.3\%.
\end{enumerate}

This study relies heavily on MsPASS and SeisBench to process raw USArray waveforms efficiently, build training and test datasets, and train and evaluate PhaseNet. \textcite{wang2022mspass} introduced MsPASS as a framework for efficient parallel processing of large seismic datasets. It provides scalable data processing and reproducible data management via NoSQL databases.
\textcite{woollam2022seisbench} described SeisBench as a framework that provides unified access to benchmark datasets, data-augmentation tools, and widely used picker models.




\section{Dataset and Workflow}
We compiled a five-year USArray Transportable Array (TA) waveform dataset (2007, 2008, 2011, 2012, and 2013) paired with manual analyst P picks from the EarthScope Array Network Facility (ANF) bulletin described by \textcite{trabant2012data}. \textcite{meltzer1999usarray} and \textcite{pavlis2010array} showed that USArray provides dense broadband coverage and supports efficient large-scale teleseismic body-wave analysis, making it a practical setting for teleseismic evaluation.


To define which raw waveforms entered the dataset, we first compiled a list of latitude-longitude pairs that we treated as source-region centers. For each year and each center, we retained earthquakes at epicentral distances of 30$^\circ$--95$^\circ$. The download included all broadband seismic channels from stations within a latitude-longitude bounding box slightly larger than the contiguous United States. In total, we downloaded 125.5 million raw USArray waveform documents. The per-year raw-waveform distribution was 3.1M in 2007, 3.2M in 2008, 112.7M in 2011, 3.6M in 2012, and 2.9M in 2013, with 2011 dominating the archive.

We then filtered the downloaded data to retain waveforms associated with at least one ANF-reported pick. Because processing 125.5 million waveforms requires an efficient pick-association workflow, we used MsPASS and its MongoDB-based indexing capabilities described by \textcite{wang2022mspass}. For clarity, we refer to each fixed-length waveform segment as a Trace, following ObsPy and SeisBench usage; in MsPASS, the same segment is stored as a document in MongoDB. We built indexes on station, channel, start time, and end time, then joined analyst picks by station and time containment ($\text{starttime} < \text{pick} < \text{endtime}$) and stored the matched pick timestamps as metadata in the documents. This join step provides a reproducible link between raw waveform segments and analyst labels at USArray scale. After associating picks with the 125.5 million raw waveforms, we selected documents with exactly one matched manual pick, yielding a final dataset of 1{,}609{,}588 waveforms in ObsPy Trace format. The per-year distribution was 200{,}719 traces in 2007, 189{,}600 in 2008, 871{,}543 in 2011, 179{,}771 in 2012, and 167{,}955 in 2013.

We then wrote the traces to per-year SeisBench HDF5/CSV files, totaling 441~GB across all years. For each year, we assigned non-overlapping train/dev/test splits of 0.8/0.1/0.1. The same splits were used throughout this study so that different model configurations were trained on the same training subset and evaluated on the same held-out test subset.

Aside from resampling to 40~Hz, we did not apply denoising or bandpass filtering to the waveforms. This choice is consistent with several modern picking studies that also use minimally processed or unfiltered waveform inputs, including the original PhaseNet formulation of \textcite{zhu2019phasenet}, regional phase-picking models reported by \textcite{aguilarsuarez2025regional} and induced-seismicity pickers developed by \textcite{heuel2025pisdl}.



\section{Models}

\begin{figure}[t]
\centering
\includegraphics[width=0.85\linewidth]{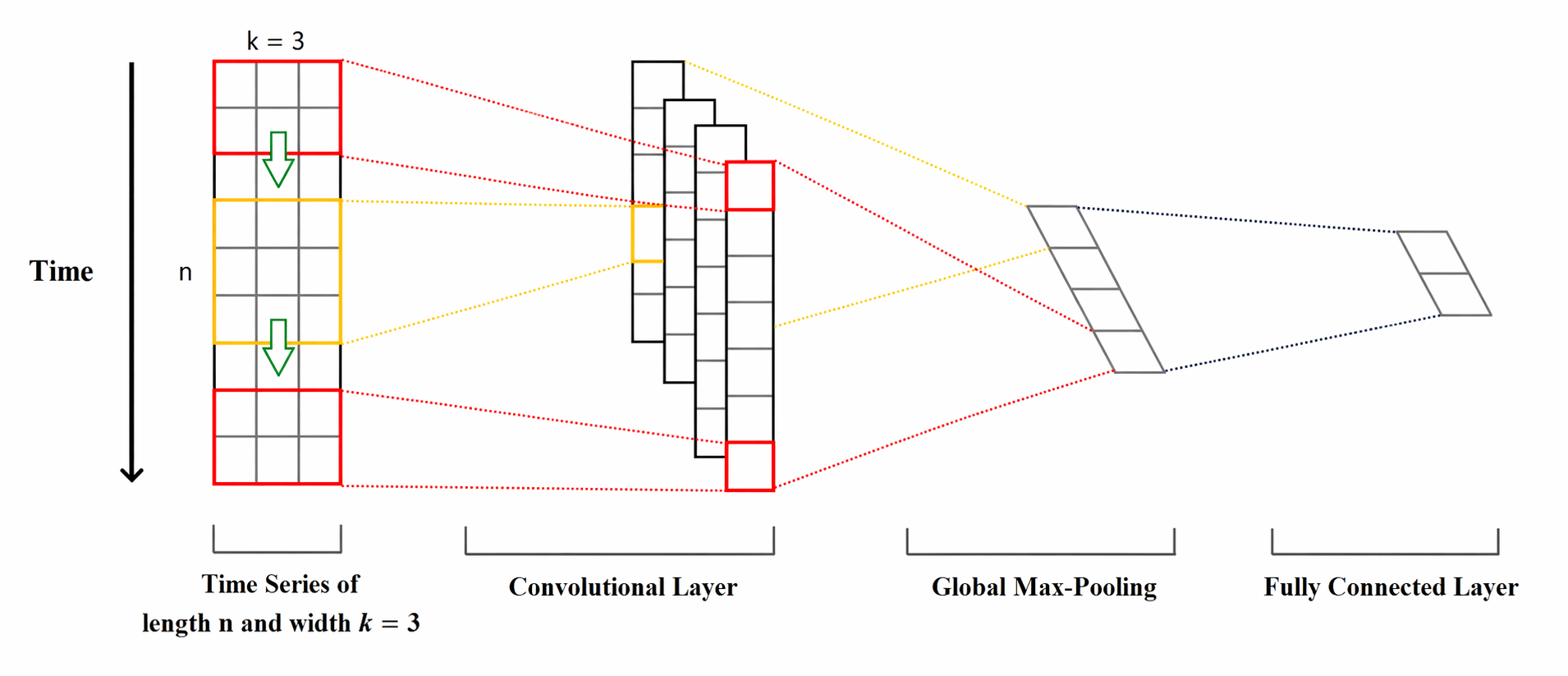}
\caption{Schematic illustration of filter-factor scaling in a generic one-dimensional convolutional neural network. The input tensor has $k=3$ waveform channels and $n=3001$ time samples. Increasing the filter factor multiplies the number of feature maps produced by each convolutional layer while preserving the temporal axis length, thereby increasing representational capacity without changing the input window. This figure is included only to clarify the channel-scaling idea used to build PhaseNet-Scale; it is not the full U-Net architecture used in PhaseNet.}
\alttext{Increasing the filter factor expands the number of channels at each convolutional stage, illustrating how PhaseNet-Scale increases model capacity without changing the waveform window.}
\label{fig:cnn}
\end{figure}

To quantify teleseismic domain shift and the impact of model-size scaling, we compare three PhaseNet configurations: (1) the standard PhaseNet model trained on the regional Northern California Earthquake Data Center (NCEDC) dataset, (2) the standard PhaseNet model trained on the USArray dataset we compiled, and (3) the same PhaseNet architecture with increased channel capacity trained on the same USArray data. We refer to these models as PhaseNet-NCEDC, PhaseNet-USArray, and PhaseNet-Scale, respectively.


Increasing the filter factor is a standard capacity-scaling strategy in convolutional networks because it expands the number of feature channels at each layer, enabling the model to represent a wider range of waveform patterns. For example, \textcite{naoi2024neural} proposed PhaseNetWC using the same idea. Figure~\ref{fig:cnn} is a simplified convolutional-neural-network schematic used only to illustrate the filter-factor concept. It is intentionally different from the U-Net architecture used in PhaseNet, whose full encoder-decoder structure is not shown here. If the input has size $3 \times 3001$, then $k = 3$ denotes the number of channels and $n = 3001$ denotes the number of samples. For example, if the original PhaseNet architecture uses 4 filters, the corresponding convolutional layer has 4 channels over 3001 samples; applying a filter factor of 8 increases the number of filters to 32. Therefore, PhaseNet-Scale uses eight times as many channels in each convolutional layer as the original PhaseNet architecture.

Table~\ref{tab:model_summary} compares the specifications of the three models. We use the PhaseNet checkpoint provided via SeisBench by \textcite{woollam2022seisbench}, which is trained on the NCEDC dataset used in the original PhaseNet paper. PhaseNet-USArray and PhaseNet-Scale are trained from scratch on USArray with randomly initialized weights. By increasing the filter factor from 1 to 8, PhaseNet-Scale reaches 32.392M parameters, making it about 120 times larger than the standard PhaseNet architecture.


\begin{table}[t]
\centering
\caption{PhaseNet configurations evaluated in this study.}
\label{tab:model_summary}
\begin{tabular}{llll}
\toprule
Model & Initialization & filter factor & Params (M) \\
\midrule
PhaseNet-NCEDC & Optimized weights obtained via SeisBench & 1 & 0.268 \\
PhaseNet-USArray & Randomly initialized & 1 & 0.268 \\
PhaseNet-Scale & Randomly initialized & 8 & 32.392 \\
\bottomrule
\end{tabular}
\end{table}

\section{Training and Evaluation}


\begin{figure}[t]
\centering
\includegraphics[width=0.70\linewidth]{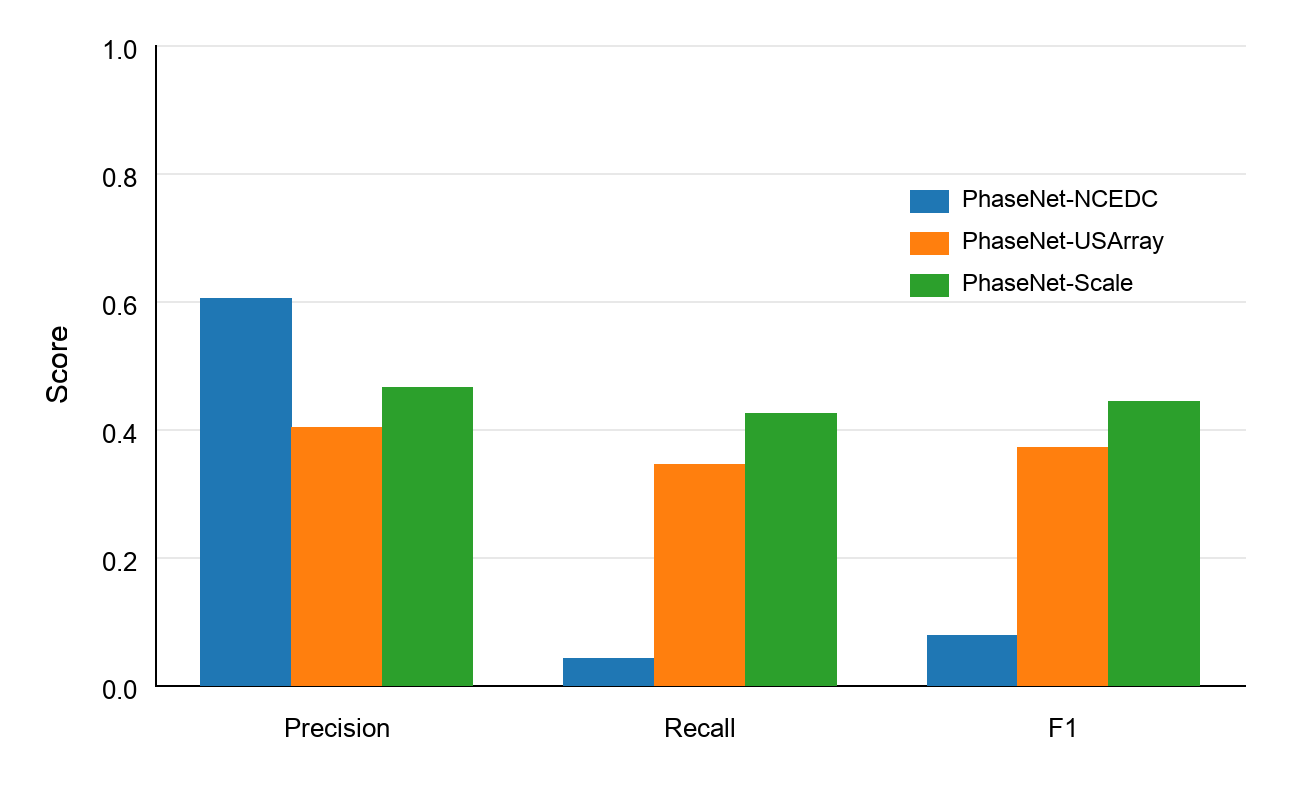}
\caption{Precision, recall, and F1 score for PhaseNet-NCEDC, PhaseNet-USArray, and PhaseNet-Scale on the held-out USArray teleseismic test set. PhaseNet-NCEDC, which was trained on regional NCEDC data, attains the highest precision but very low recall, indicating severe under-triggering under teleseismic domain shift. Training the standard architecture on USArray data substantially improves recall and F1, and the larger PhaseNet-Scale model provides an additional but smaller gain over PhaseNet-USArray.}
\alttext{Regional-data training yields very low recall on teleseismic records. Retraining on USArray greatly increases recall and F1, and the larger PhaseNet-Scale model gives the best overall balance of the three metrics.}
\label{fig:prf}
\end{figure}

\begin{figure}[t]
\centering
\includegraphics[width=0.85\linewidth]{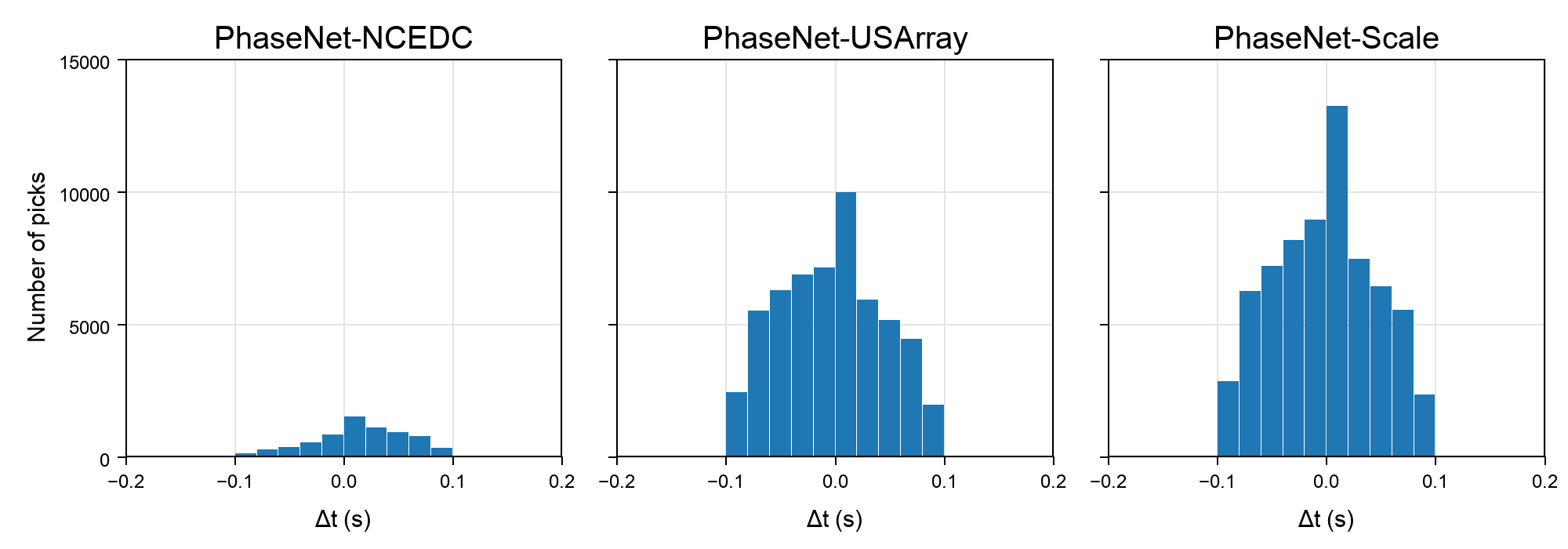}
\caption{P-pick residual histograms for PhaseNet-NCEDC, PhaseNet-USArray, and PhaseNet-Scale on the held-out USArray teleseismic test set. Residuals are defined as predicted pick time minus analyst pick time, and all panels use the same binning and y-axis scale over the plotted $\pm 0.1$~s range so that counts can be compared directly across models. Relative to PhaseNet-NCEDC, the USArray-trained models produce many more picks concentrated near zero residual, with PhaseNet-Scale showing the strongest central concentration and therefore the largest number of highly accurate picks.}
\alttext{The USArray-trained models produce far more picks near zero residual than the regional baseline, with PhaseNet-Scale showing the strongest clustering around zero and therefore the most accurate analyst-aligned picks.}
\label{fig:residual_hist}
\end{figure}

\begin{table}[t]
\centering
\caption{P-pick accuracy on USArray teleseismic test data.}
\label{tab:pick_metrics}
\begin{tabular}{lcccc}
\toprule
Model & Precision & Recall & F1 & Picks within $\pm 0.1$~s \\
\midrule
PhaseNet-NCEDC & 0.603 & 0.041 & 0.077 & 4{,}564 \\
PhaseNet-USArray & 0.403 & 0.345 & 0.372 & 35{,}774 \\
PhaseNet-Scale & 0.466 & 0.425 & 0.444 & 44{,}840 \\
\bottomrule
\end{tabular}
\end{table}

We use the 3-channel windows as input to PhaseNet and focus on evaluating predicted P picks against analyst P picks. Each example in the dataset is a single-component ObsPy Trace containing the waveform component associated with the analyst P pick. For compatibility with PhaseNet, each trace is converted into a 3-channel input window by placing the observed P-waveform component in one channel and zero-padding the other two channels. We optimize a cross-entropy loss between predicted
phase probabilities and probabilistic labels:
\begin{equation}
\mathcal{L} = -\sum_{t} \sum_{c \in \{P,S,N\}} y_{t,c}\log p_{t,c},
\end{equation}
where $p_{t,c}$ is the predicted probability of class $c$ at sample $t$ and $y_{t,c}$ is the probabilistic
label. We optimize the model parameters using the Adam optimizer with a learning rate of 1e-4 and a batch size of 512. Early stopping is applied based on the dev set: the model is evaluated every 5 training epochs, and training is terminated if the dev loss does not improve for two consecutive validation checks.

After training, we evaluate the models on the held-out test split from all years. We declare a P pick if the maximum P-class probability over the window satisfies
$\max_t p_P(t) \ge 0.5$. For precision, recall, and F1, we count a predicted pick as correct when the predicted and manual pick times differ by no more than 0.1~s. We additionally report the number of picks within the same 0.1~s residual window.

Precision, recall, and F1 are standard metrics for evaluating prediction performance, and we use them here to compare model behavior. Precision measures the fraction of predicted picks that are correct, recall measures the fraction of analyst picks recovered by the model, and F1 summarizes the balance between precision and recall. Specifically, we compute
\begin{eqnarray}
\text{Precision} = \frac{TP}{TP + FP}   \\
\text{Recall} = \frac{TP}{TP + FN} \\
\text{F1} = \frac{2PR}{P + R} 
\end{eqnarray}
where $TP$ is true positives, $FP$ is false positives, $FN$ is false negatives, $P$ is precision, and $R$ is recall. 

Overall performance improves dramatically from PhaseNet-NCEDC to PhaseNet-USArray, while PhaseNet-Scale provides a smaller additional gain over PhaseNet-USArray. Table~\ref{tab:pick_metrics} and Figure~\ref{fig:prf} summarize these differences. Specifically,

\begin{itemize}
\item PhaseNet-NCEDC under-triggers on teleseismic data, with the highest precision (0.603) but the lowest recall (0.041), so many true P arrivals are missed.
\item Relative to PhaseNet-NCEDC, PhaseNet-USArray increases recall by 741.5\% (0.041 to 0.345), F1 by 383.1\% (0.077 to 0.372), and the number of picks within $\pm 0.1$~s by 683.9\% (4{,}564 to 35{,}774), while precision decreases by 33.2\% (0.603 to 0.403).
\item Relative to PhaseNet-USArray, PhaseNet-Scale provides a smaller but consistent improvement, increasing precision by 15.6\% (0.403 to 0.466), recall by 23.2\% (0.345 to 0.425), and F1 by 19.4\% (0.372 to 0.444), while raising the number of picks within $\pm 0.1$~s by 25.3\% (35{,}774 to 44{,}840).
\end{itemize}

Residual histograms (Figure~\ref{fig:residual_hist}) visualize the distribution of pick residuals over the plotted $\pm 0.1$~s range. Compared with PhaseNet-NCEDC, PhaseNet-USArray produces many more picks near zero residual, with most residuals concentrated within about $\pm 0.05$~s. PhaseNet-Scale shows the strongest concentration near zero, indicating the largest number of highly accurate picks. In addition to the 19.4\% gain in F1, these histograms suggest that PhaseNet-Scale also improves pick-time accuracy.

Overall, PhaseNet-Scale delivers the best performance, but PhaseNet-USArray captures most of the accuracy gains with only 0.268M parameters, whereas PhaseNet-Scale expands to 32.392M parameters (about 120 times larger) for a smaller incremental improvement.

\section{Benchmark}


\begin{table}[t]
\centering
\caption{Benchmark on CPU and GPU.}
\label{tab:benchmark}
\resizebox{\linewidth}{!}{%
\begin{tabular}{lrrrr}
\toprule
Model & Params (M) & Checkpoint (MiB) & Infer thr. GPU (samples/s) & Infer thr. CPU (samples/s)\\
\midrule
PhaseNet-USArray & 0.268 & 1.06 & $99778.96$ & $1100.54$ \\
PhaseNet-Scale & 32.392 & 123.65 & $12799.27$ & $29.22$\\
\bottomrule
\end{tabular}
}
\end{table}

\begin{figure}[t]
\centering
\includegraphics[width=0.95\linewidth]{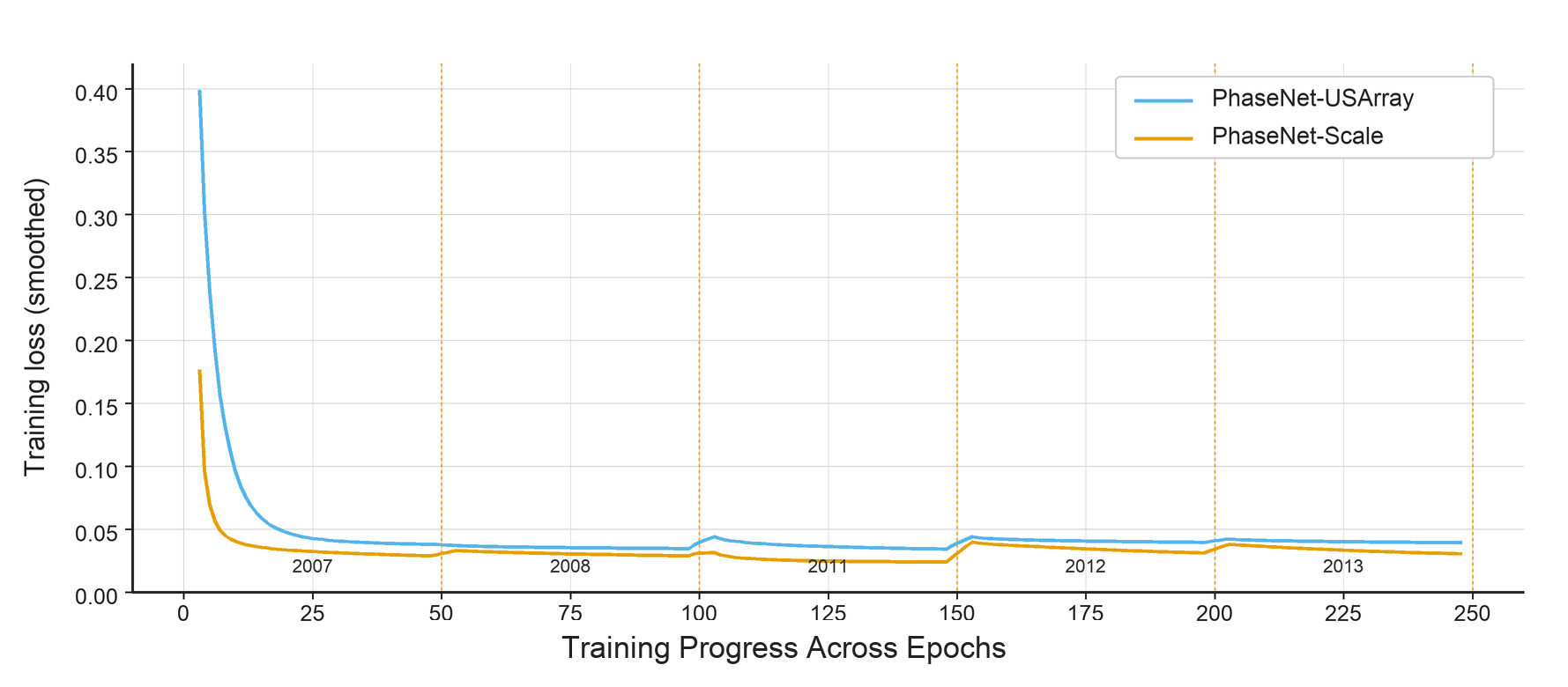}
\caption{Training loss curves for PhaseNet-USArray and PhaseNet-Scale across sequential year-by-year training on the USArray dataset. Both models reach low loss values early in training, indicating stable optimization under the same Adam-based training setup. The larger PhaseNet-Scale model converges faster and reaches a slightly lower minimum loss, while the small upward steps at year boundaries reflect transitions between yearly data files rather than training instability.}
\alttext{Both models reach low loss quickly during year-by-year training. PhaseNet-Scale drops faster and ends slightly lower, while small jumps at year boundaries are brief and followed by renewed decline.}
\label{fig:train_loss}
\end{figure}


This section quantifies the efficiency of PhaseNet-USArray and PhaseNet-Scale using training loss and inference throughput. We benchmark training and inference on a single NVIDIA A100 GPU. In addition, CPU inference runs were executed on a single Lonestar6 compute node (one node with dual AMD EPYC 7763 64-core CPUs, 128 cores total, 256~GB RAM, and 288~GB local \texttt{/tmp} SSD).


As shown in Figure~\ref{fig:train_loss}, both PhaseNet-USArray and PhaseNet-Scale converge to a low training loss ($<0.05$) within the first 25 epochs, while PhaseNet-Scale converges even faster, within the first 10 epochs. Because the TACC Lonestar \texttt{/tmp} filesystem has a 288~GB limit, and we use \texttt{/tmp} staging to reduce I/O overhead, we train sequentially year by year rather than loading all years at once. At year transitions, the loss shows only a small temporary increase and then drops again within the first few epochs, often reaching values equal to or lower than those from the previous year.

Overall, PhaseNet-USArray and PhaseNet-Scale show very similar training-loss trends, while PhaseNet-Scale reaches a slightly lower minimum loss. This observation is consistent with the use of the same Adam optimizer policy for both models and with the greater capacity of the larger model, which can fit the training set more effectively.

The inference-throughput evaluation focuses solely on model inference speed and excludes other steps in the inference pipeline, such as I/O and data loading, batch construction, and host-to-device transfer in GPU runs. We estimate throughput by running each model's best checkpoint on the evaluation set 10 times and reporting the mean number of samples processed per second. As listed in Table~\ref{tab:benchmark}, we observed:
\begin{itemize}
    \item On the GPU, PhaseNet-Scale achieves 12.8\% of the throughput of PhaseNet-USArray (12{,}799.27 versus 99{,}778.96 samples/s).
    \item On the CPU, PhaseNet-Scale achieves only 2.7\% of the throughput of PhaseNet-USArray (29.22 versus 1{,}100.54 samples/s).
\end{itemize}


With 120 times more parameters (from $0.268\,\mathrm{M}$ to $32.392\,\mathrm{M}$), PhaseNet-Scale improves recall and F1 by $23.2\%$ and $19.4\%$, respectively, but reduces inference throughput by $87.2\%$ on GPU and $97.3\%$ on CPU. This comparison leads to two conclusions:
\begin{enumerate}
    \item Scaling PhaseNet is more practical on GPUs than on CPUs.
    \item The accuracy gain obtained by scaling PhaseNet is not proportional to the efficiency loss; for example, a $23.2\%$ gain in recall comes with an $87.2\%$ reduction in GPU throughput. Therefore, simply enlarging PhaseNet is not an efficient way to achieve large accuracy gains.
\end{enumerate}

\section{Conclusions}
We presented a reproducible MsPASS-SeisBench workflow, compiled a USArray teleseismic dataset with 1{,}609{,}588 ready-to-use waveforms in ObsPy Trace format, and evaluated PhaseNet under domain shift. PhaseNet trained on a regional dataset shows high precision but extremely low recall (precision = 0.603, recall = 0.041, F1 = 0.077), making it unsuitable for teleseismic deployment. Training PhaseNet from scratch on USArray restores sensitivity (precision = 0.403, recall = 0.345, F1 = 0.372), increasing F1 by 383.1\% and yielding 683.9\% more P picks (35{,}774) within $\pm 0.1$~s of analyst picks.

The PhaseNet-Scale variant, which is 120 times larger than PhaseNet-USArray, further improves accuracy (precision = 0.466, recall = 0.425, F1 = 0.444), increasing F1 by 19.4\% and generating 25.3\% more P picks than PhaseNet-USArray. However, PhaseNet-Scale reduces inference throughput by 87.2\% on GPU and 97.3\% on CPU.

Together, these results show that a model trained only on regional data does not transfer reliably to teleseismic records. Training PhaseNet on teleseismic data recovers many previously missed arrivals relative to PhaseNet-NCEDC, and a larger model can provide additional gains, but at a substantial efficiency cost.

\section{Data and Resources}
Waveform data and analyst picks used in this study were obtained from the USArray Transportable Array archive documented by \textcite{doi.org/10.7914/sn/ta} and the EarthScope Array Network Facility bulletin reported by \textcite{trabant2012data}. Open-source software used in the workflow includes MsPASS, presented by \textcite{wang2022mspass}; SeisBench, introduced by \textcite{woollam2022seisbench}; and PhaseNet, introduced by \textcite{zhu2019phasenet}.

To support reproducibility, we made the necessary materials available at \url{https://github.com/mspass-team/mspass_tutorial/tree/master/usarray_phasenet_eval} (last accessed May 2026). They include a small subset of the compiled USArray dataset, the PhaseNet-USArray model checkpoint, and the data-processing, training, and evaluation code. The full dataset is not included because of its size (441 GB), but it is available from the authors upon reasonable request.

Researchers using teleseismic P phases should find the workflows in this paper a useful starting point for assembling a working dataset. Notable examples include teleseismic P-wave tomography and P-wave receiver-function imaging. With the transfer of the EarthScope archives to a cloud data provider, it should soon be feasible to adapt this workflow to estimate a consistent set of teleseismic P-wave picks from the entire EarthScope archive. We also note that this work would not have been feasible without the MsPASS framework for processing a dataset of this scale (125.5 million waveforms).

\section{Acknowledgments}
The authors acknowledge the Texas Advanced Computing Center (TACC) at The University of Texas at Austin for providing computational resources that have contributed to the research results reported within this paper.

\printbibliography

@article{allen1978automatic,
  title={Automatic earthquake recognition and timing from single traces},
  author={Allen, Rex V},
  journal={Bulletin of the seismological society of America},
  shortjournal={Bull. Seism. Soc. Am.},
  volume={68},
  number={5},
  pages={1521--1532},
  year={1978},
  publisher={The Seismological Society of America}
}

@article{saragiotis2002pai,
  title={PAI-S/K: A robust automatic seismic P phase arrival identification scheme},
  author={Saragiotis, Christos D and Hadjileontiadis, Leontios J and Panas, Stavros M},
  journal={IEEE transactions on geoscience and remote sensing},
  shortjournal={IEEE Trans. Geosci. Remote Sens.},
  volume={40},
  number={6},
  pages={1395--1404},
  year={2002},
  publisher={IEEE}
}

@article{zhu2019phasenet,
  title={PhaseNet: a deep-neural-network-based seismic arrival-time picking method},
  author={Zhu, Weiqiang and Beroza, Gregory C},
  journal={Geophysical Journal International},
  shortjournal={Geophys. J. Int.},
  volume={216},
  number={1},
  pages={261--273},
  year={2019},
  publisher={Oxford University Press}
}

@article{ross2018generalized,
  title={Generalized seismic phase detection with deep learning},
  author={Ross, Zachary E and Meier, Men-Andrin and Hauksson, Egill and Heaton, Thomas H},
  journal={Bulletin of the Seismological Society of America},
  shortjournal={Bull. Seism. Soc. Am.},
  volume={108},
  number={5A},
  pages={2894--2901},
  year={2018},
  publisher={GeoScienceWorld}
}

@article{mousavi2020earthquake,
  title={Earthquake transformer—an attentive deep-learning model for simultaneous earthquake detection and phase picking},
  author={Mousavi, S Mostafa and Ellsworth, William L and Zhu, Weiqiang and Chuang, Lindsay Y and Beroza, Gregory C},
  journal={Nature communications},
  shortjournal={Nat. Commun.},
  volume={11},
  number={1},
  pages={3952},
  year={2020},
  publisher={Nature Publishing Group UK London}
}

@article{soto2021deepphasepick,
  title={DeepPhasePick: A method for detecting and picking seismic phases from local earthquakes based on highly optimized convolutional and recurrent deep neural networks},
  author={Soto, Hugo and Schurr, Bernd},
  journal={Geophysical Journal International},
  shortjournal={Geophys. J. Int.},
  volume={227},
  number={2},
  pages={1268--1294},
  year={2021},
  publisher={Oxford University Press}
}

@article{munchmeyer2022picker,
  title={Which picker fits my data? A quantitative evaluation of deep learning based seismic pickers},
  author={M{\"u}nchmeyer, Jannes and Woollam, Jack and Rietbrock, Andreas and Tilmann, Frederik and Lange, Dietrich and Bornstein, Thomas and Diehl, Tobias and Giunchi, Carlo and Haslinger, Florian and Jozinovi{\'c}, Dario and others},
  journal={Journal of Geophysical Research: Solid Earth},
  shortjournal={J. Geophys. Res.: Solid Earth},
  volume={127},
  number={1},
  pages={e2021JB023499},
  year={2022},
  publisher={Wiley Online Library}
}

@article{wang2022mspass,
  title={MsPASS: A data management and processing framework for seismology},
  author={Wang, Yinzhi and Pavlis, Gary L and Yang, Weiming and Ma, Jinxin},
  journal={Seismological Research Letters},
  shortjournal={Seismol. Res. Lett.},
  volume={93},
  number={1},
  pages={426--434},
  year={2022}
}

@article{woollam2022seisbench,
  title={SeisBench—A toolbox for machine learning in seismology},
  author={Woollam, Jack and M{\"u}nchmeyer, Jannes and Tilmann, Frederik and Rietbrock, Andreas and Lange, Dietrich and Bornstein, Thomas and Diehl, Tobias and Giunchi, Carlo and Haslinger, Florian and Jozinovi{\'c}, Dario and others},
  journal={Seismological Research Letters},
  shortjournal={Seismol. Res. Lett.},
  volume={93},
  number={3},
  pages={1695--1709},
  year={2022}
}

@article{naoi2024neural,
  title={Neural phase picker trained on the Japan meteorological agency unified earthquake catalog},
  author={Naoi, Makoto and Tamaribuchi, Koji and Shimojo, Kengo and Katoh, Shinya and Ohyanagi, Shukei},
  journal={Earth, Planets and Space},
  shortjournal={Earth Planets Space},
  volume={76},
  number={1},
  pages={150},
  year={2024},
  publisher={Springer}
}

@article{yu2023benchmark,
  title={Benchmark on the accuracy and efficiency of several neural network based phase pickers using datasets from China Seismic Network},
  author={Yu, Ziye and Wang, Weitao and Chen, Yini},
  journal={Earthquake Science},
  shortjournal={Earthq. Sci.},
  volume={36},
  number={2},
  pages={113--131},
  year={2023},
  publisher={Elsevier}
}

@article{meltzer1999usarray,
  title={The usarray initiative},
  author={Meltzer, Anne and Rudnick, Roberta and Zeitler, Peter and Levander, Alan and Humphreys, Gene and Karlstrom, Karl and Ekstrom, E and Carlson, C and Dixon, Tim and Gurnis, Michael and others},
  journal={Geological Society of America Today},
  shortjournal={GSA Today},
  volume={9},
  pages={8--10},
  year={1999},
  publisher={Geological Society of America}
}

@article{pavlis2010array,
  title={Array processing of teleseismic body waves with the USArray},
  author={Pavlis, Gary L and Vernon, Frank L},
  journal={Computers \& Geosciences},
  shortjournal={Comput. \& Geosci.},
  volume={36},
  number={7},
  pages={910--920},
  year={2010},
  publisher={Elsevier}
}

@article{aguilarsuarez2025regional,
  title={Picking Regional Seismic Phase Arrival Times with Deep Learning},
  author={Aguilar-Suarez, Carlos and Beroza, Gregory C},
  journal={Seismica},
  shortjournal={Seismica},
  volume={4},
  number={1},
  year={2025},
  doi={10.26443/seismica.v4i1.1431}
}

@article{heuel2025pisdl,
  title={Picking Induced Seismicity with Deep Learning (piSDL)},
  author={Heuel, Nils and Quinteros-Cartaya, Carolina and Benites, Marcos and Kraft, Toni and Plenkers, Katrin and Sippl, Christian and Bindi, Dino and Cesca, Simone},
  journal={Seismica},
  shortjournal={Seismica},
  volume={4},
  number={2},
  year={2025},
  doi={10.26443/seismica.v4i2.1579}
}

@article{johnson1979cedar,
  title={CEDAR: An approach to the computer automation of short-period local seismic networks, 1. Seismotectonics of the Imperial Valley of southern California},
  author={Johnson, Carl Edward},
  journal={Ph. D. Thesis},
  year={1979}
}

@article{kaplan2020scaling,
  title={Scaling laws for neural language models},
  author={Kaplan, Jared and McCandlish, Sam and Henighan, Tom and Brown, Tom B and Chess, Benjamin and Child, Rewon and Gray, Scott and Radford, Alec and Wu, Jeffrey and Amodei, Dario},
  journal={arXiv preprint arXiv:2001.08361},
  year={2020}
}

@misc{doi.org/10.7914/sn/ta,
  doi = {10.7914/SN/TA},
  url = {https://www.fdsn.org/networks/detail/TA/},
  author = {{IRIS Transportable Array}},
  title = {USArray Transportable Array},
  publisher = {International Federation of Digital Seismograph Networks},
  year = {2003}
}

@article{trabant2012data,
  title={Data products at the IRIS DMC: Stepping stones for research and other applications},
  author={Trabant, Chad and Hutko, Alexander R and Bahavar, Manochehr and Karstens, Richard and Ahern, Timothy and Aster, Richard},
  journal={Seismological Research Letters},
  shortjournal={Seismol. Res. Lett.},
  volume={83},
  number={5},
  pages={846--854},
  year={2012},
  publisher={Seismological Society of America}
}

\end{document}